\documentclass[pre,showpacs,twocolumn]{revtex4}
\usepackage{amsfonts}
\usepackage{amsmath}
\usepackage[dvips]{graphicx}

\renewcommand{\vec}[1]{\mathbf{ #1 }}

\newcommand{\myint}[1]{ \int\limits_{\mathbb{R}^{n}} \! d^{n}\vec{#1}\, }

\begin{document}

\title{Eigenvalue cut-off in the cubic-quintic nonlinear Schr\"odinger equation}
\author{Vladyslav Prytula$^1$}
\author{Vadym Vekslerchik$^{1,2}$} 
\author{V\'\i ctor M. P\'erez-Garc\'\i a$^1$} 

\affiliation{$^1$Departamento de Matem\'aticas, E.T.S. Ingenieros Industriales and  Instituto de Matem\'atica Aplicada a la Ciencia y la Ingenier\'{\i}a, Universidad de Castilla-La Mancha, Avda. Camilo Jos\'e Cela 3, Ciudad Real, 13071 Spain
\\
$^2$A. Ya. Usikov Institute for Radiophysics and Electronics \\
National Academy of Sciences of Ukraine, \\
12 Proskura Street, 61085 Kharkov, Ukraine
}

\begin{abstract}
Using theoretical arguments, we prove the numerically well-known fact that the eigenvalues of all localized stationary solutions of the cubic-quintic 2D+1 nonlinear Schr\"odinger equation exhibit an upper cut-off value. The existence of the cut-off is inferred using Gagliardo-Nirenberg and H\"older inequalities together with Pohozaev identities. We also show that, in the limit of eigenvalues close to zero, the eigenstates of the cubic-quintic nonlinear Schr\"odinger equation behave similarly to those of the cubic nonlinear Schr\"odinger equation.   
\end{abstract}

\pacs{ 05.45.Yv, 42.65.Tg}

\date{\today}
\maketitle

\section{Introduction}

The nonlinear Schr\"odinger  equation (NLSE) has been widely used in modeling nonlinear wave dynamics in many physical scenarios, such as nonlinear optics \cite{Kivshar,Hasegawa},  plasma physics \cite{Dodd}, Bose-Einstein condensates \cite{Dalfovo}, biomolecular dynamics \cite{Davidov}, and others \cite{Sulem,PG}. The simplest scalar NLSE is of the form
\begin{equation}
i\frac{\partial \psi}{\partial t} = -\Delta \psi + f\left(|\psi|^2\right)\psi.
\end{equation}
where $\psi$ is a complex field defined usually on the whole $\mathbb{R}^n$ with $n=1,2,3$ and $f$ describes the nonlinear response of the medium. 
Many different types of nonlinearities $f(|\psi|^2)$ arise in the different physical fields of applicability of the equation including: power-law, saturable and nonlocal nonlinearities to cite a few examples. The most relevant one, for which many  theoretical and analytical studies of NLS equations have been done is the classical cubic nonlinearity $f(|\psi|^2) = g |\psi|^2$, both because of its direct interest and also because it corresponds to the simplest nonlinear response proportional to the square of the involved field (namely the light intensity in optics, the number of particles in BEC applications, etc...).

One of the simplest extensions of the cubic NLS is the so-called cubic-quintic (CQNLS) model, which in normalized units, is
\begin{equation}\label{Cqui}
i\frac{\partial \psi}{\partial t} = -\Delta \psi - g|\psi|^2\psi+h|\psi|^4\psi.
\end{equation}
The CQNLS equation is another universal mathematical model describing many situations of physical interest and approximating other more complicated ones. As examples it arises in plasma physics \cite{CQplasma,Stab6}, condensed matter physics \cite{CQcm}, nuclear physics \cite{CQnuclear}, Bose-Einstein condensation \cite{CQBEC}, etc., but probably the application of the model which has attracted more attention in the last years is description of the propagation of paraxial beams in certain nonlinear optical media. Many different optical materials have a refractive index that can be well described by a cubic-quintic nonlinearity such as some semiconductors and doped glasses (e.g.  AlGaAs \cite{AlGas} and CdS$_x$Se$_{1-x}$ \cite{Cd}), 
the polydiacetylene para-toluene sulfonate (PTS) \cite{PTS}, chalcogenide glasses \cite{Chal}, some transparent organic materials \cite{trans} or even media with complex susceptibilities induced by Electromagnetically Induced Transparency \cite{Pazlight}.

We will consider localized stationary solutions (i.e. solitary waves or solitons) of Eq. (\ref{Cqui}). Thus,  taking $\psi(x,t) = u(x) e^{i\beta t}$ we will study solutions of
\begin{equation}\label{Cqui2}
\beta u = \Delta u + \left(g|u|^2-h|u|^4\right)u.
\end{equation}
with $u$ a complex function defined on $\mathbb{R}^2$ and vanishing at infinity. 

The  stationary solutions of Eq. (\ref{Cqui2}) and their stability properties, have been studied in many papers \cite{Stab1,ref:liquid_light,Stab2,ref:dimitrievski,Yang,Stab3,ref:berezhiani,ref:jovanovski,Liquidlight,Stab4,Stab5,ref:pego,Stab6,Stab7,ref:humberto_vortex,Davydova,Stab72,Stab75}. 

A well-known fact is that no solitary waves are found beyond a maximum value of the eigenvalue $\beta  = \beta_*$. In fact for $\beta \leq \beta_*$ one finds many solutions with different widths corresponding to very different values of the norm of the solution $N_{\beta} = \int_{\mathbb{R}^2} |u_{\beta}|^2 dx$. In fact, when $\beta \rightarrow \beta_*$, $N_{\beta} \rightarrow \infty$. Thus, there is a cut-off in the permitted eigenvalue of  localized solutions of the CQNLS. 
The existence of this cutoff and its specific value as a function of the parameters was discussed in many papers. For fundamental states it was studied numerically \cite{anderson} and by approximate variational methods with a supergaussian ansatz in Ref. \cite{ref:dimitrievski} and also in Ref. \cite{Liquidlight}. For vortex states the same problem was studied by means of variational \cite{ref:humberto_vortex}, numerical \cite{ref:humberto_vortex} and analytical methods \cite{ref:pego,ref:berezhiani}.

In this paper we provide a theoretical support for  previous works dealing with the problem of the cutoff \cite{ref:dimitrievski,Liquidlight,ref:humberto_vortex,ref:pego,ref:berezhiani} and prove in a rigorous yet simple way that 
 localized stationary states of the cubic-quintic NLS equation exist only for eigenvalues on a finite interval. The obtained result holds for all type of localized solutions, e.g. ground, vortex or dipole states, of the cubic-quintic NLS equation. 
 
The article is organized as follows: first in Section \ref{sec-statement} we introduce the NLS equation to be considered, the integral quantities and the inequalities to be used in the subsequent sections. Section \ref{sec-cutoff} contains the derivation of the upper limit for the eigenvalues and analysis of the behavior near cut--off. Next, in Section \ref{sec-wide} we describe the related problem of the asymptotic behavior of the stationary states in the NLSE limit. Finally, in Sec. \ref{sec-conclusion}  we summarize our results.

\section{Statement of the problem}

\label{sec-statement}

The typical situation for the cubic-quintic nonlinearity corresponds to the case where we have a combination of focusing cubic and defocusing quintic term. The interplay of focusing and defocusing nonlinearity in that case prevents the wave collapse \cite{Zakharov} and is responsible for the liquid-like features of the localized stationary states \cite{ref:liquid_light,Liquidlight,ref:humberto_vortex}. 

In this section we will consider the case of two spatial dimensions which is the one relevant to nonlinear optics and many other applications of the model.

The equation we are dealing with can be written as
\begin{equation}
\beta u =  \Delta u + g u^3 - hu^5\:, \label{eq:cq_focusing}
\end{equation}
with $\beta,g,h>0.$  Our analysis will use the energy
\begin{equation}
K  = 
- \beta N +g U_2 - h U_3
\label{energy}
\end{equation}
and  Pohozaev identities \cite{ref:Pohozaev, ref:Lions-Berestycki}
\begin{eqnarray}
0 = \beta N -\frac{g}{2} U_2 + \frac{h}{3} U_3,
\label{pohozaev}
\end{eqnarray}
where the moments of  $u$ are defined by:
\begin{subequations}
\label{eq:moments}
\begin{eqnarray}
K & = & \myint{r} |\nabla u|^2,
\\
N &=& \myint{r} u^{2},
\\
U_j & = & \myint{r} u^{2j}, \quad j=2,3. 
\end{eqnarray}
\end{subequations}

\section{Cutoff in the cubic-quintic NLSE}
\label{sec-cutoff}
\subsection{Existence of cut-off}

By direct calculations from (\ref{energy}) and (\ref{pohozaev}) we get the following \textit{bilinear} relation between the moments
\begin{multline}
4hK^2 +\left(3g^2-16\beta h\right)N K + \beta\left(16\beta h-3g^2\right)N^2 \\
+g^2 h \left(NU_3-U_2^2\right)=0.
\end{multline}
This relation can be rewritten in terms of $X = N/K$
\begin{equation}
\beta\left(\beta -\beta_* \right)X^2+\left(\beta_*-\beta\right)X+\frac{1}{4}\left(1+\Gamma\right)=0,
\label{quadratic-eq}
\end{equation}
where
\begin{equation}
\Gamma= \frac{g^2}{4}\,
\frac{NU_3-U_2^2}{K^2},\qquad 
\beta_* = \frac{3}{16}\,\frac{g^2}{h}.
\end{equation}
It should be noted that it follows from the H\"older inequality,
 \begin{equation}
NU_3 \geq U_2^2,
\end{equation} 
that
$
\Gamma>0
$.
Now, since our quadratic equation (\ref{quadratic-eq}) must have real roots, after calculating its discriminant 
\begin{equation}
\mathcal{D} = \left(\beta_*-\beta\right)\left(\beta_* +\beta\Gamma\right)
\end{equation}
one can conclude that
\begin{equation}
\beta< \beta_*.
\label{bound_beta}
\end{equation}

Thus, there is an upper bound on the eigenvalues, $3g^2/16h$ and  localized stationary states of Eq. (\ref{eq:cq_focusing}) exist only when $\beta$ lies within 
the interval 
\begin{equation}
\beta\in \left(0,\frac{3}{16}\frac{g^2}{h}\right).
\end{equation}
This result agrees with previous numerical and approximate calculations for 
this quantity \cite{ref:dimitrievski,Liquidlight,ref:humberto_vortex,ref:pego,ref:berezhiani} but here is obtained rigorously using simple arguments. In fact, the comparison with the numerical results of Ref. \cite{ref:dimitrievski,Liquidlight,ref:humberto_vortex} shows that our bound is optimal and matches closely the numerics.

\subsection{Behavior near cut-off}
\label{rescaling}
A natural question that arises, after establishing the bound (\ref{bound_beta}) is what occurs with solution when the parameter $\beta$ approaches the cut-off limit. The aim of this section is to obtain a bound for the quantity $N$ which allows us to show that $N\to +\infty$ when $\beta\to \beta_*$.

Solving Eq. (\ref{quadratic-eq}) we obtain that
\begin{equation}
\frac{N}{K} = \frac{1}{2\beta}
\left(
	1+
	\sqrt{\frac{\beta_*+\Gamma\beta}{\beta_*-\beta}}
	\label{roots}
\right).
\end{equation}
The sign before the radical is determined by the identity
\begin{equation}
  2\beta N - K =\frac{1}{2}g U_2 >0,
  \label{U2}
\end{equation}
which is a direct consequence of Eqs. (\ref{energy}) and (\ref{pohozaev}).
Since $\Gamma>0$, Eq. (\ref{roots}) leads to the inequality
\begin{equation}
\frac{N}{K} > \frac{1}{2\beta}\;
	\sqrt{\frac{\beta_*}{\beta_*-\beta}}.
	\label{roots_bound}
\end{equation}
On the other hand we can obtain another type of bounds for $K$ and $N$. From Eqs. (\ref{energy}) and  (\ref{pohozaev}) we get
\begin{subequations}
\begin{eqnarray}
\frac{h}{3} U_3 & = & \beta N - K,
\label{N-K} \\
K & = & \frac{1}{2}gU_2- \frac{2}{3}hU_3.
\label{K_U_2_U3}
\end{eqnarray}
\end{subequations}
Combining these relations and the identity 
\begin{equation}
\frac{\partial K}{\partial \beta} = N,
\label{derivative_K}
\end{equation}
which can be derived by differentiating Eq. (\ref{eq:cq_focusing}) and applying again Eqs.  (\ref{energy}) and (\ref{pohozaev}), 
we conclude that $K$ is a monotone function. These leads to the fact that
\begin{equation}
K = 0\quad \mbox{when}\quad \beta=0
\end{equation}
and respectively $K = \int \limits_{0}^{\beta}N \, d\beta$.
Now let us recall the particular form of Gagliardo-Nirenberg \cite{gagliardo, nirenberg,ref:PinoDolbeaut} inequality which in our case can be written as
\begin{equation}
U_2\leq C_{\scriptscriptstyle GN} K N,
\end{equation}
where $C_{\scriptscriptstyle GN}$ is the optimal constant for the Gagliardo-Nirenberg inequality in two dimensions. Applying this inequality to Eq. (\ref{K_U_2_U3}) we get
\begin{equation}\label{NN}
N \geq \frac{2}{C_{\scriptscriptstyle GN}g}.
\end{equation}
Combining Eqs. (\ref{NN}) and  (\ref{derivative_K}) we get the inequality
\begin{equation}
K \geq \frac{2}{C_{\scriptscriptstyle GN}g}\beta.
\end{equation}
Substituting this estimate into (\ref{roots_bound}) we  obtain  finally
\begin{equation}\label{Nbeta}
N \geq \frac{1}{C_{\scriptscriptstyle GN}}\;
	\sqrt{\frac{\beta_*}{\beta_*-\beta}},
\end{equation}
which guarantees that $N\to +\infty$ as $\quad \beta\to \beta_*\equiv 3g^2/(16h)$.
Thus we can see that the cut-off phenomenon is a manifestation of the blow-up of the norm (note that $N$ is nothing but the $L^2$-norm of $u$).

Another consequence of the estimate (\ref{Nbeta}) is the fact that $N$ is bounded from below by $1/C_{\scriptscriptstyle GN}$ when $\beta \to \, 0$.

\section{The limit $\beta\rightarrow 0$}
\label{sec-wide}

It is posible to analyze the asymptotic behavior of solutions of Eq. (\ref{eq:cq_focusing})  when $\beta\rightarrow 0$. To do it we use the rescaling symmetry of our equation, specifically we will use the fact that if 
$u$ is a solution of Eq. (\ref{eq:cq_focusing}) then the function $\tilde{u}$ given by 
$\tilde{u}(\vec{r})= \xi u(\eta\vec{r})$ solves 
\begin{equation}
\tilde{\beta}u = \Delta \tilde{u} +\tilde{g}\tilde{u}^3-\tilde{h}\tilde{u}^5,
\label{eq:scaled_eq}
\end{equation}
where 
\begin{equation}
\tilde{\beta}=\eta^2\beta, \quad \tilde{g}=\frac{\eta^2}{\xi^2}g,\quad \tilde{h} = \frac{\eta^2}{\xi^4}h.
\end{equation}
Calculating the moments of the function $\tilde{u}$ we arrive at the following rescaling relation for $K$ and $N$ treated as  functions of the parameters $\beta$, $g$ and $h$:
\begin{subequations}
\label{eq:moments_rescaled}
\begin{eqnarray}
K (\beta, g, h)& = & \frac{1}{\xi^2}K\left(\eta^2\beta,\frac{\eta^2}{\xi^2}g,\frac{\eta^2}{\xi^4}h\right),
\\
N (\beta, g, h)& = & \frac{\eta^2}{\xi^2}N\left(\eta^2\beta,\frac{\eta^2}{\xi^2}g,\frac{\eta^2}{\xi^4}h\right),
\end{eqnarray}
\end{subequations}
Further, choosing 
$$\eta^2=\frac{1}{\beta},\quad\xi^2=\frac{g}{\beta}$$  
we can obtain
\begin{subequations}
\label{eq:moments_rescaled_res}
\begin{eqnarray}
K(\beta,g,h)&=& \frac{\beta}{g} K\left(1,1, \frac{\beta h}{g^2}\right),
\\[3mm]
N (\beta, g, h)& = & \frac{1}{g}N\left(1,1,\frac{\beta h}{g^2}\right),
\end{eqnarray}
\end{subequations}
Considering the limit $\beta\to 0$ we find the following asymptotic formulae for $N[u]$ and $K[u]$:
\begin{subequations}
\begin{eqnarray}
N[u]  & = & \frac{1}{g}N[\phi] + O(\beta) \\
K[u]  & =&  \frac{\beta}{g}K[\phi] + 
       \frac{\beta^2 h}{3g^3}U_3[\phi]+O(\beta^3),
\end{eqnarray}
\end{subequations}
where $\phi$ is the Townes soliton, i.e. the solution of 
\begin{equation}
  \Delta \phi = \phi - \phi^3.
\end{equation}
 
These results make it possible to deduce that in the asymptotic region as $\beta\to 0$ the behavior of solutions of cubic-quintic Schr\"odinger equation is similar to the behavior of eigensolutions of the cubic Nonlinear Schr\"odinger equation. 

\section{Conclusions}
\label{sec-conclusion}

Using general mathematical arguments we have shown that the localized stationary solutions of the cubic-quintic Nonlinear Schr\"odinger equation exhibit an eigenvalue cut-off. This result holds for all type of localized stationary solutions, e.g. fundamental, vortex or dipole states. The cut--off  is in excellent agreement with previous approximate or numerical results. 
We have also obtained a lower bound for the number of particles in the  eigenstates that diverges exactly at the cutoff $\beta = \beta_*$. Finally, in the limit of small eigenvalues we obtain the result that  the eigenstates of the cubic quintic nonlinear Schr\"odinger equation behave in the same way as those of the cubic nonlinear Schr\"odinger equation. 

Our results complement present knowledge on one of the key models of mathematical physics and support in a rigorous yet simple way previous numerical observations. 
 
\acknowledgments
This work has been partially supported by Ministerio de Educaci\'on y Ciencia (MEC), Spain (project  FIS2006-04190) and Junta de Comunidades de Castilla-La Mancha (project PCI-08-0093). V. P.  acknowledges support from MEC through Grant No. AP2005-4528.


\begin{thebibliography}{09}

\bibitem{Kivshar}{Y. Kivshar, G. P. Agrawal, {\em Optical Solitons: From fibers to Photonic crystals}, Academic Press (2003).}

\bibitem{Hasegawa}{A. Hasegawa, {\em Optical Solitons in Fibers},
 Springer-Verlag, Berlin, (1989).}

\bibitem{Dodd}{R.K. Dodd, J.C. Eilbeck, J.D. Gibbon, H.C. Morris,
{\em Solitons and nonlinear wave equations}, Academic Press, New
York (1982).}

\bibitem{Dalfovo}{F. Dalfovo, S. Giorgini , L. P. Pitaevskii , S. Stringari,
 {\em Theory of Bose-Einstein condensation in trapped gases}, Rev. Mod. Phys. 71:463-512 (1999).}

\bibitem{Davidov}{A.S. Davydov, {\em Solitons in Molecular Systems}
Reidel, Dordrecht (1985).}

\bibitem{Sulem}
C. Sulem and P. Sulem, 
{\it The Nonlinear Schr\"odinger Equation: Self-focusing and Wave Collapse} (Springer, Berlin, 2000).

\bibitem{PG} L. V{\'a}zquez, L. Streit, V. M. P{\'e}rez-Garc{\'\i}a,
      Eds., \emph{Nonlinear Klein-Gordon and Schr{\"o}dinger systems: Theory and
      Applications}, World Scientific, Singapur (1997).

\bibitem{CQclassic} Kh. I. Pushkarov, D. I. Pushkarov, and I. V. Tomov, Opt.
Quantum Electron. \textbf{11}, 471 (1979); Kh. I. Pushkarov and D. I.
Pushkarov, Rep. Math. Phys. \textbf{17}, 37 (1980).

\bibitem{CQplasma} C. T. Zhou and X. T. He, Phys. Scr. \textbf{50} 415 (1994).

\bibitem{Stab6} T. A. Davydova, A. I. Yakimenko and Yu. A. Zaliznyak, Phys. Rev. E \textbf{67}, 026402 (2003).

\bibitem{CQcm} I. V. Barashenkov and V. G. Makhankov V G, Phys. Lett. A \textbf{28}
52 (1988); I. V. Barashenkov and E. Yu Panova Physica D \textbf{69} 114 (1993).

\bibitem{CQnuclear} V. G. Kartavenko, J. Nucl. Phys. \textbf{40} 377 (1984).

\bibitem{CQBEC} A. Bulgac, Phys. Rev. Lett. \textbf{89}, 050402 (2002); F. Kh. Abdullaev, A. Gammal, L. Tomio, and T. Frederico,
Phys. Rev. A \textbf{63}, 043604 (2001); .A.-X. Zhang and J.-K. Xue 
Phys. Rev. A \textbf{75} 013624 (2007)

 \bibitem{AlGas} S. Tanev and D. I. Pushkarov, Opt. Commun. \textbf{141}, 322 (1997).
 
 \bibitem{Cd} L. H. Acioli, A. S. L. Gomes, J. M. Hickmann, and C. B. de
Araujo, Appl. Phys. Lett. \textbf{56}, 2279 (1990); P. Roussignol, D. Ricard, J. Lukasik, and C. Flytzanis, J. Opt.
Soc. Am. B \textbf{4}, 5 (1987).

\bibitem{PTS} B. L. Lawrence and G. I. Stegeman, Opt. Lett. \textbf{23}, 591(1998).

 \bibitem{Chal} K. Ogusu, J. Yamasaki, S. Maeda, M. Kitao, and M. Minakata,
Opt. Lett. 29, 265 (2004).

 \bibitem{trans} C. Zhan, D. Zhang, D. Zhu, D. Wang, Y. Li, Z. Lu, L. Zhao,
and Y. Nie, J. Opt. Soc. Am. B 19, 369 (2002).
 
\bibitem{Pazlight} H. Michinel, M. J. Paz-Alonso, and V. M. Perez-Garcia, Phys. 
Rev. Lett. \textbf{96}, 023903 (2006).

\bibitem{Stab1} M. Quiroga-Teixeiro, H. Michinel, J. Opt. Sot. Am. B \textbf{14}, 2004 (1997).

\bibitem{ref:liquid_light} C. Josserand and S. Rica, Phys. Rev. Lett. {\bf 78}, 1215 (1997).

\bibitem{Stab2}  D. V. Skryabin and W. J. Firth Phys. Rev. E \textbf{58}, 3916 (1998).

\bibitem{ref:dimitrievski}
K. Dimitrevski, E. Reimhult, E. Svensson, A. \"Ohgren, D. Anderson, A. Berntson, M. Lisak and M. L. Quiroga-Teixeiro, 
Phys. Lett. A {\bf 248}, 369 (1998).

\bibitem{Yang} J. Yang, D. J. Kaup, SIAM J. Appl. Math. \textbf{60},967 (2000).

\bibitem{Stab3} I. Towers, A. V. Buryak, R. A. Sammut, B. A. Malomed, L.-C. Crasovan and D. Mihalache, Phys. Lett. A \textbf{288}, 292 (2001).

\bibitem{ref:berezhiani}
V. I. Berezhiani, V. Skarka, and N. B. Aleksi\'c,
Phys. Rev. E {\bf 64}, 057601 (2001).

\bibitem{ref:jovanovski}
Z. Jovanovski, 
J. Mod. Opt. {\bf 48}, 865-875 (2001).

\bibitem{Liquidlight} H. Michinel, J. Campo-Taboas, R. Garc\'{\i}a-Fern\'andez, J. R. Salgueiro, and M. L. Quiroga-Teixeiro, Phys. Rev. E {\bf 65}, 066604 (2002).

\bibitem{Stab4} B. A. Malomed, L. C. Crasovan and D. Mihalache, Physica D \textbf{161}, 187 (2002).

\bibitem{Stab5} D. Mihalache, D. Mazilu, L.-C. Crasovan, I. Towers, A. V. Buryak, B. A. Malomed, L. Torner, J. P. Torres, and F. Lederer, Phys. Rev. Lett. 88, 073902 (2002).

\bibitem{ref:pego}
R.L. Pego and H.A. Warchall, 
J. Nonlinear Sci. {\bf 12}, 347 (2002).

\bibitem{Stab7} D. Mihalache, D. Mazilu, I. Towers, B. A. Malomed and F. Lederer, Phys. Rev. E \textbf{67} 056608 (2003).

\bibitem{ref:humberto_vortex} H. Michinel, J. R. Salgueiro and M. J. Paz-Alonso, 
Phys. Rev. E {\bf 70}, 066605 (2004).

\bibitem{Davydova} T. A. Davydova, A. I. Yakimenko, J. Opt. A: Pure Appl. Opt. \textbf{6}, S197 (2004).

\bibitem{Stab72} H. Michinel, J. R. Salgueiro, and M. J. Paz-Alonso, 
Phys. Rev. E \textbf{70}, 066605 (2004).

\bibitem{Stab75} A. S. Desyatnikov, D. Mihalache, D. Mazilu, B. A. Malomed, C. Denz, and F. Lederer
Phys. Rev. E \textbf{71}, 026615 (2005).

\bibitem{anderson} D. Anderson, G. Derrick, J. Math. Phys. \textbf{11}, 1336-1346 (1970); \textbf{12}, 945-952 (1971).

\bibitem{Stab8} D. Mihalache, D. Mazilu, F. Lederer, Y. V. Kartashov, L.-C. Crasovan, L. Torner, and B. A. Malomed, 
Phys. Rev. Lett. \textbf{97}, 073904 (2006).

\bibitem{Zakharov} N.E. Kosmatov, V.F. Shvets, V.E. Zaharov, Physica D \textbf{52}, 19 (1991).

\bibitem{ref:Pohozaev}
S. I. Pohozaev, 
Sov. Math. Doklady {\bf 5}, 1408-1411 (1965).

\bibitem{ref:Lions-Berestycki}
H. Berestycki and P.L. Lions,  
Arch. for Rational Mechanics  and Analysis {\bf 82}, 313 (1983).

\bibitem{ref:Strauss}
W. A. Strauss, 
Comm. Math. Phys. {\bf 55}, 149-162 (1977).

\bibitem{ref:PinoDolbeaut}
M. Del Pino and J. Dolbeault, 
Jour. Math. Pur. Appliq. {\bf 9}, 847  (2002).

\bibitem{ref:Cazenave}
T. Cazenave, 
{\it Introduction to Nonlinear Schr\"odinger Equations } (Textos de M\'etodos Matem\'aticos, Rio de Janeiro, 1989).

\bibitem{ref:Adams}
R. Adams,
{\it Sobolev Spaces } (Academic Press, 1975).

\bibitem{nirenberg}
L. Nirenberg,  Ann. Sc. Norm. Pisa {\bf 13}, 116--162
(1959).

\bibitem{gagliardo}
E. Gagliardo, Ric. Mat. {\bf 7}, 102--137 ,(1958).

\end{thebibliography}
\end{document}